\documentclass[preprint,3p,11pt]{elsarticle}
\biboptions{compress}

\usepackage{psfrag}
\usepackage{axodraw4j}
\usepackage{pstricks}
\usepackage{color}
\usepackage{hyperref} 
\usepackage{amsmath}
\usepackage{epsfig}
\usepackage{amsfonts}
\usepackage{amssymb}
\usepackage{booktabs,multirow,tabularx}
\usepackage{multirow}
\usepackage{bbold}

\newcommand{\amc}{{\sc\small MadGraph5\_aMC@NLO}}
\newcommand{\madspin}{{\sc\small MadSpin}}
\newcommand{\mg}{{\sc\small MadGraph5}}
\newcommand{\mfks}{{\sc\small MadFKS}}
\newcommand{\ml}{{\sc\small MadLoop}}
\newcommand{\fj}{{\sc\small FastJet}}
\newcommand{\ct}{{\sc\small CutTools}}
\newcommand{\ol}{{OpenLoops}}
\newcommand{\sushi}{{\sc\small SusHi}}

\begin{document}
\begin{flushright}
ZU-TH 22/14\\
\end{flushright}
\vskip1cm

\title{Rare Standard Model processes for present and future hadronic colliders}
\author{Paolo Torrielli}
\address{Physik-Institut, Universit\"at Z\"urich, Winterthurerstrasse 190, 8057 Z\"urich, Switzerland}

\begin{abstract}
\begin{small}
In this talk\footnote{Talk presented at the 1st Future Hadron Collider Workshop, CERN, May 26-28, 2014.} I present the total cross sections, accurate at the NLO in QCD, for rare Standard-Model hadroproduction processes involving multi-Higgs-boson, multi-electroweak-boson and multi-top-quark final states. The comparison between cross sections at the LHC and at a future circular hadronic collider with up to 100 TeV centre-of-mass energy is detailed.
Results relevant to the hadronic production of five electroweak bosons, and of a top-antitop pair in association with an electroweak vector boson and two jets are presented here for the first time with NLO accuracy.
\end{small}
\end{abstract}
\maketitle

\section{Introduction}
As the LHC is approaching its Run-II phase, and the experimental collaborations are rapidly progressing in the assessment and reduction of the systematic uncertainties affecting their measurements, on the theoretical side it is necessary to deliver reliable and accurate predictions for as complex reactions as possible, such as ones featuring many electroweak vector bosons, Higgs bosons, and top quarks in the final state. The availability of these predictions is crucial both in searches for New Physics, in order to have all sorts of Standard-Model (SM) backgrounds under very good theoretical control, and for the determination of the characteristics and quantum numbers of the recently discovered Higgs boson \cite{Aad:2012tfa,Chatrchyan:2012ufa}, where rare signals may in some cases represent a particularly clean detection environment. Moreover, with a wider perspective, some of the processes that can be considered rare and of challenging difficulty for the LHC setup, may instead turn out to be very well visible at the future circular hadronic collider (FCC-hh), due to the foreseen increase in luminosity and centre-of-mass energy.

It is thus important for this kind of studies, entailing the simulation of quite diverse hadronic reactions at various collider energies, to employ robust and flexible tools, best if based on a unique and consistent computational framework, and with no need of resorting to different codes for different types of processes. Moreover, in the case of a detailed analysis of multi-particle final states, NLO QCD accuracy can turn out to be mandatory for a reliable prediction of both the normalisation and the shape of differential distributions and for the assessment of the associated theoretical uncertainties, and several classes of observables require a logarithmic resummation, like the one performed by parton showers, for a complete coverage of the phase space.

The public program \amc~\cite{Alwall:2014hca} is a framework that combines all of the above-mentioned elements. It allows the user to automatically simulate generic SM reactions at the NLO in QCD, matched to parton showers, starting from a minimal amount of input information (e.g. type of process required and value of parameters). Its generality stems from the fact that the elements specific to the particular physics process considered are generated dynamically on the basis of the physical input, and not hard-coded beforehand. The simulations of this environment are based on the tree-level-diagram generation of \mg~\cite{Alwall:2011uj}, on the NLO FKS~\cite{Frixione:1995ms} infrared subtraction implemented in \mfks~\cite{Frederix:2009yq}, and on the loop-diagram generation of \ml~\cite{Hirschi:2011pa}, that in turn relies on the OPP method \cite{Ossola:2006us} implemented in \ct~\cite{Ossola:2007ax}, and is optimised through the \ol~\cite{Cascioli:2011va} technique (although the program allows to interface to third-party one-loop providers). Its NLO matching with parton showers is achieved by means of the MC@NLO formalism \cite{Frixione:2002ik}, applied to all modern collinear-based Monte Carlo's \cite{Frixione:2002ik,Torrielli:2010aw,Frixione:2010ra,PY8unp}, and its NLO multi-jet merging is based on the FxFx procedure \cite{Frederix:2012ps}.

In what follows I presents results, all obtained with \amc, relative to the total cross section for various rare SM reactions at the LHC and at the FCC-hh. Many of the cross sections reported here are already known in the literature at the NLO in QCD for LHC energies (see for example section 4.1 of \cite{Alwall:2014hca}, and references therein), and have been recomputed here with the sole aim of extracting, readily and in a consistent setup, the cross-section increase (denoted as $\rho$ in the next sections) at the 100 TeV FCC-hh with respect to the 8 TeV LHC. Results for the hadronic production of five electroweak vector bosons, and of a top-antitop pair in association with an electroweak vector boson and two jets are instead presented here for the first time at the NLO.

This talk is structured as follows: in section \ref{sec:setup}, I describe the physical setup used
to generate all results; sections \ref{sec:H}, \ref{sec:V}, and \ref{sec:t} present predictions for processes involving multi-Higgs-boson, multi-electroweak-bosons, and multi-top-quark final states, respectively; in section \ref{sec:conc}, I draw my conclusions.  

\section{Setup}
\label{sec:setup}
In this section I summarise the physical setup employed to obtain the results shown below. \begin{itemize}
\item Non-zero particle masses are $m_t=173$ GeV, $m_H=125$ GeV, $m_Z=91.188$ GeV, $m_W=80.419$ GeV. The bottom-quark mass is set to $m_b=4.7$ GeV in the four-flavour-scheme (4FS) simulations, and to $m_b=0$ in the five-flavour-scheme (5FS) ones. The CKM matrix is $V_{\tiny{\mbox{CKM}}}=\mathbb1$, and the fine-structure constant is $\alpha=1/132.507$.
\item Renormalisation and factorisation scales are chosen as $\mu_R=\mu_F=\frac12\sum_k m_T^{(k)}$, $m_T^{(k)}$ being the transverse mass of the $k$-th final-state particle. Independent variation of $\mu_R$ and $\mu_F$ in the range $[1/2,2]$ is obtained in an exact way without rerunning the code, through the reweighting technique described in \cite{Frederix:2011ss}. The uncertainty associated with this variation is shown as a dark band in the plots of the following sections.
\item As parton distribution functions (PDFs) I use the MSTW 2008 NLO (68\%~c.l.) sets \cite{Martin:2009iq}, relevant to four or five active flavours, depending on the flavour scheme employed in the simulation. PDF uncertainties are estimated according to the asymmetric-hessian prescription provided by the PDF set, and obtained automatically as in explained in \cite{Frederix:2011ss}. They are shown as a light band in the plots of the following sections. The value and the running of the strong coupling constant $\alpha_{\tiny{\mbox{S}}}$ are as well set according to the PDF set.
\item Whenever relevant, jets are clustered by means of the anti-$k_T$ algorithm \cite{Cacciari:2008gp} as implemented in \fj~\cite{Cacciari:2011ma}, with $\Delta R=0.7$, and $p_T(j)>20$ GeV (if not otherwise specified).
\item Whenever relevant, photons are isolated by means of the Frixione smooth-cone criterion \cite{Frixione:1998jh}, with parameters $R_0=0.4$, $p_T(\gamma)>20$ GeV, $\epsilon_\gamma=n=1$.
\end{itemize}

\section{Multi-Higgs production}
\label{sec:H}
\begin{figure}[h!]
\begin{minipage}{0.49\textwidth}
\centering
\includegraphics[width=1\textwidth]{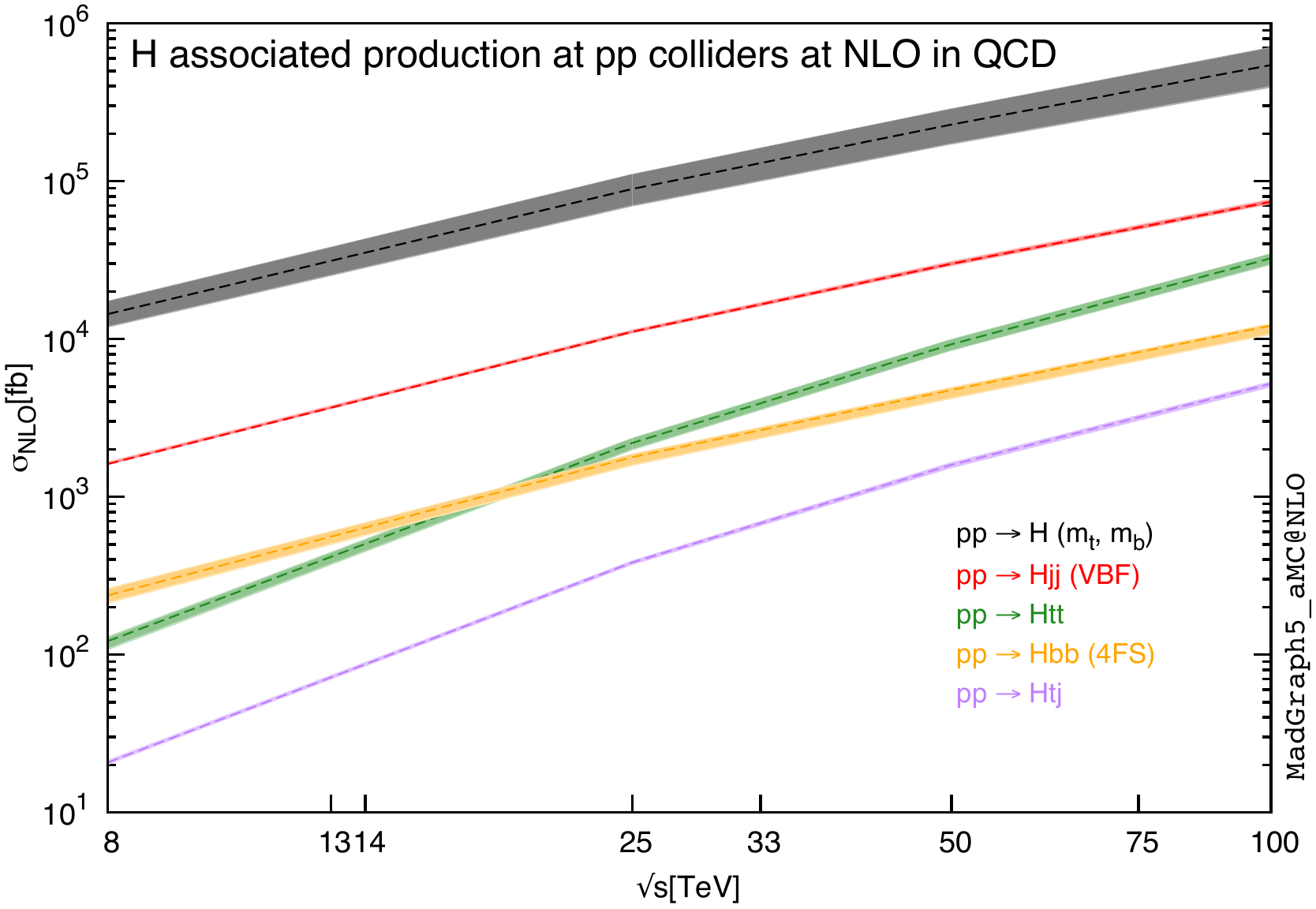}
\end{minipage}
\begin{minipage}{0.49\textwidth}
\centering
\includegraphics[width=1.02\textwidth]{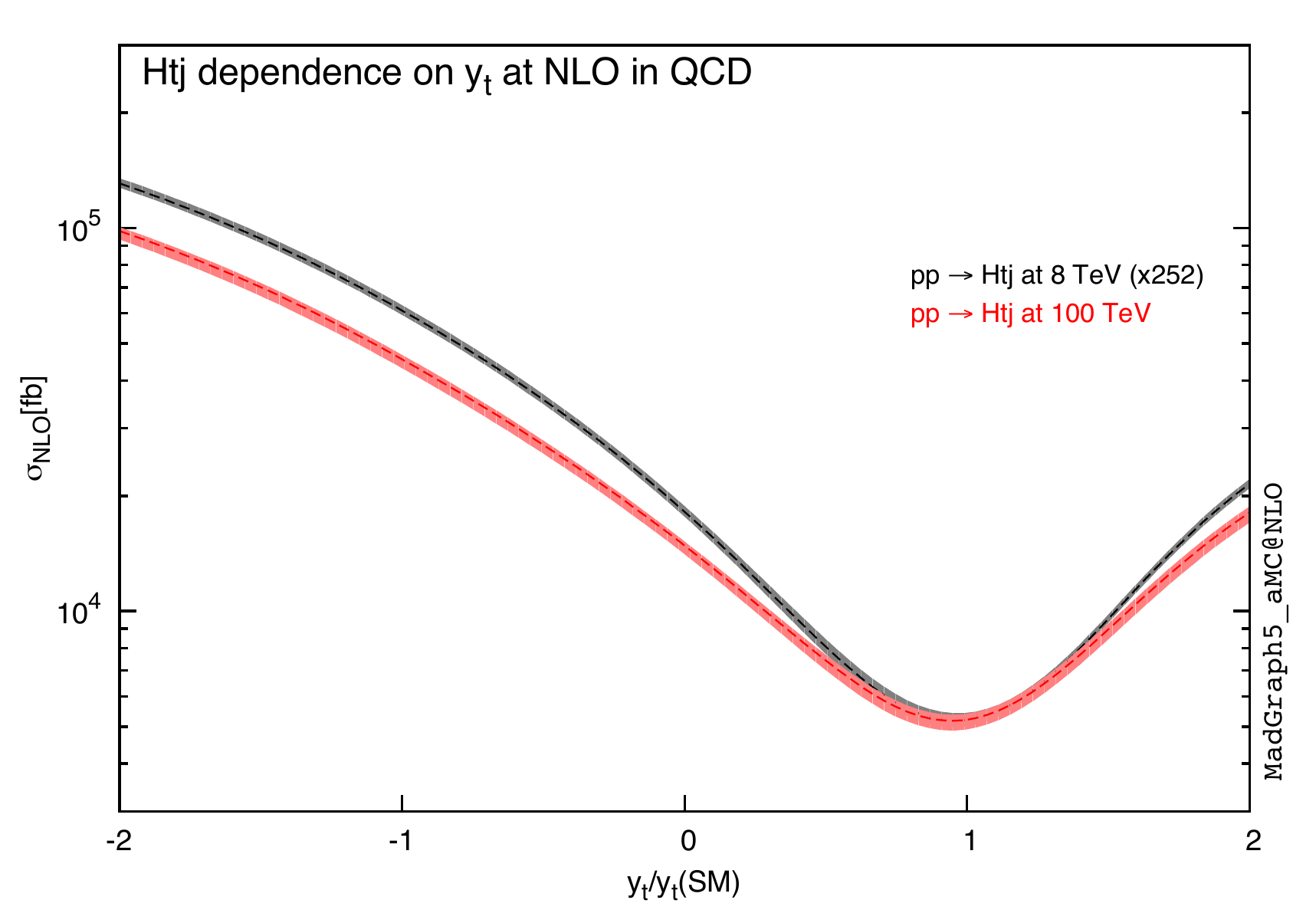}
\end{minipage}
\caption{\label{fig:H}Left panel: NLO total cross section for Higgs production in association with light and heavy quarks. Right panel: dependence of the $pp\to Htj$ NLO total cross section on the top-quark Yukawa coupling $y_t$. The 8 TeV LHC result is rescaled up by a factor 252 (see table \ref{tab:H}) in order for its SM cross section to coincide with the FCC-hh one.}
\end{figure}

\begin{figure}[h!]
\begin{minipage}{0.49\textwidth}
\centering
\includegraphics[width=1.01\textwidth]{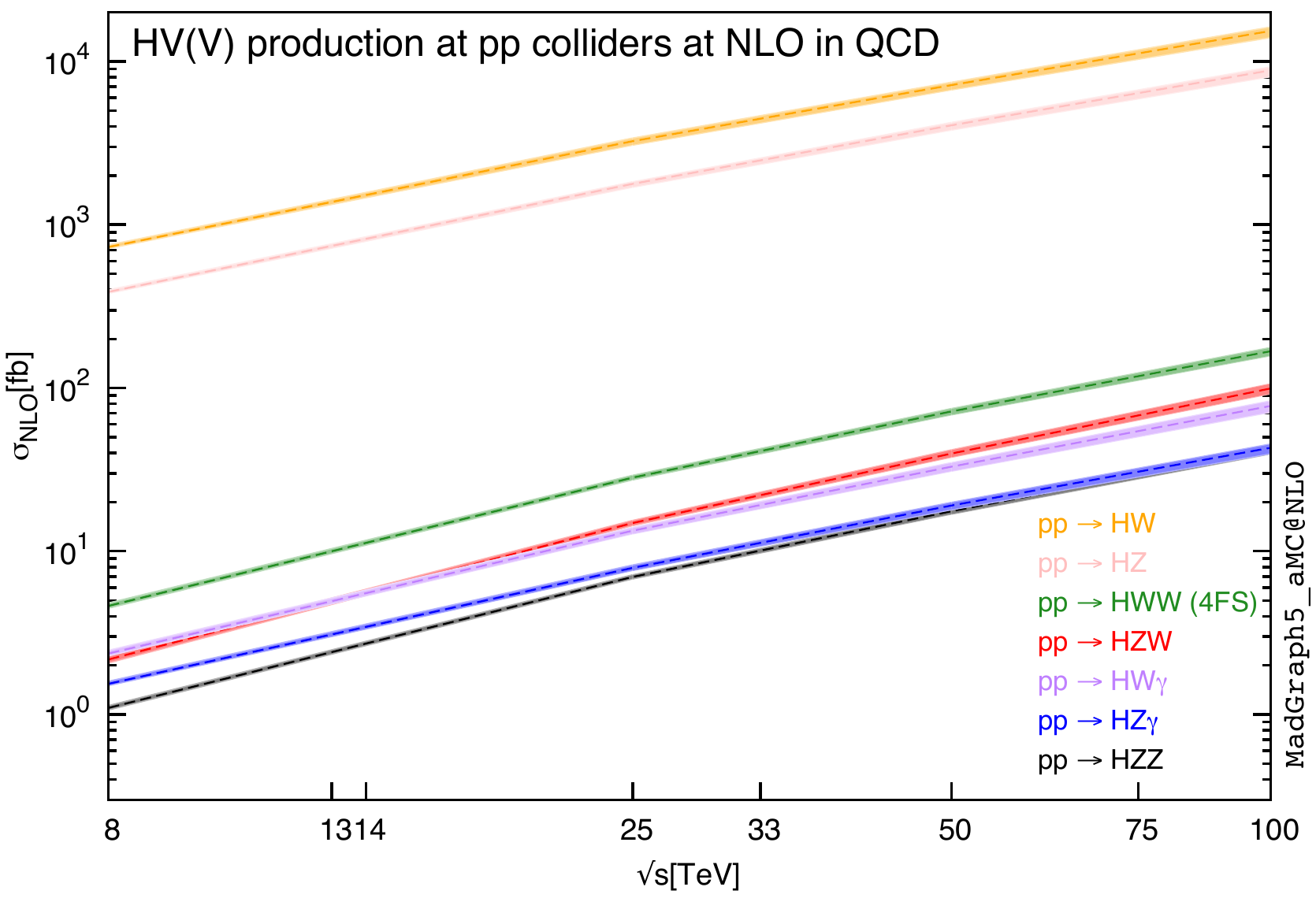}
\end{minipage}
\begin{minipage}{0.49\textwidth}
\centering
\includegraphics[width=1\textwidth]{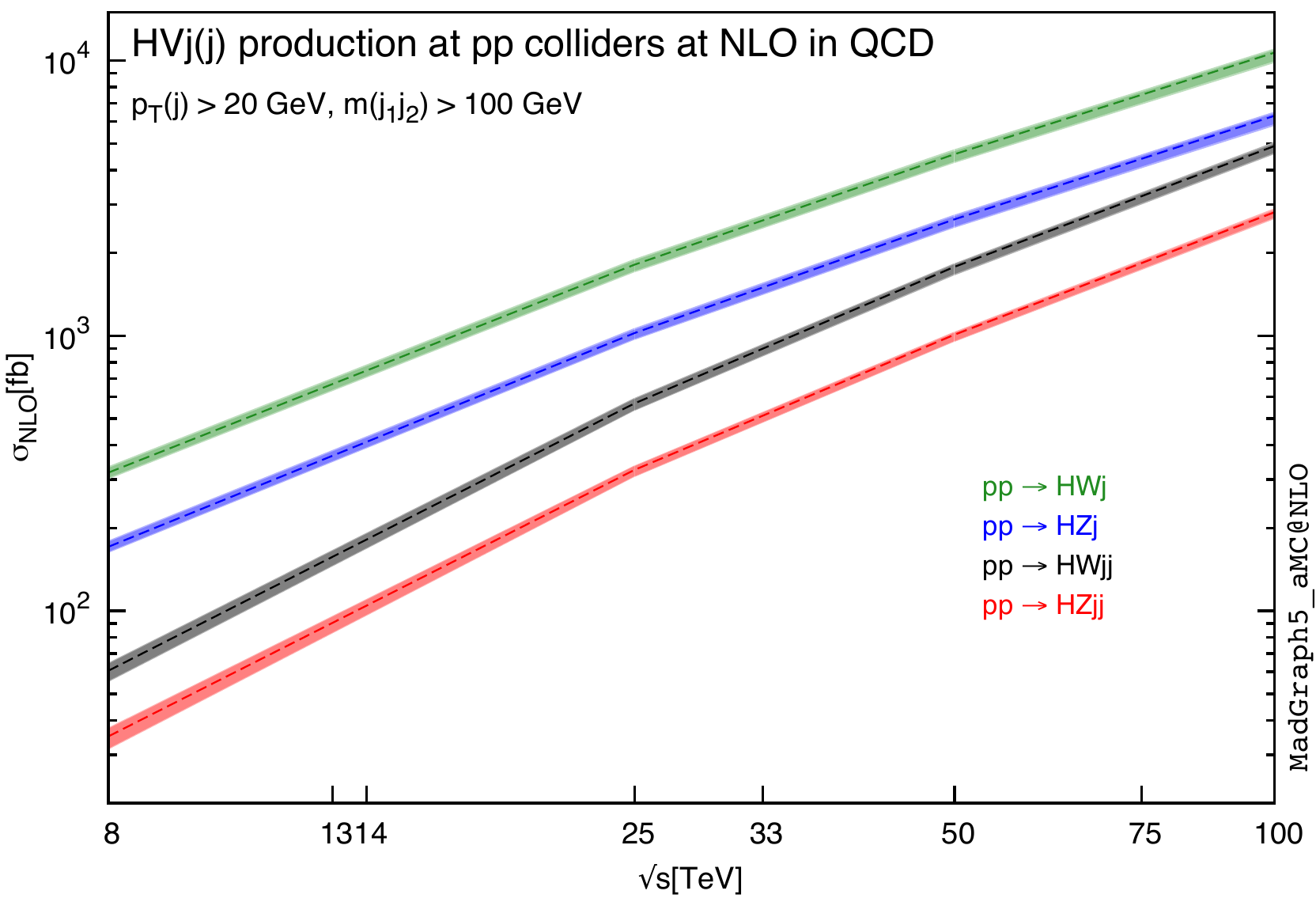}
\end{minipage}
\caption{\label{fig:HV}NLO total cross section for Higgs production in association with up to two electroweak bosons (left panel), and with an electroweak boson and up to two jets (right panel). Cuts on the jet system are described in section \ref{sec:setup} and in the caption of table \ref{tab:H}.}
\end{figure}

In table \ref{tab:H} and in figures \ref{fig:H} (left panel) and \ref{fig:HV}, I display the NLO total cross section for the production of a single Higgs boson in association with jets, heavy quarks and electroweak bosons, as a function of the collider centre-of-mass energy. Higgs production through gluon fusion, $pp\to H~(m_t, m_b)$, includes exact top- and bottom-mass dependence, though in a non-automated way, through an interface to the program \sushi~\cite{Harlander:2012pb} realised by H. Mantler and M. Wiesemann (in a way similar to what done for analytic transverse-momentum resummation in \cite{Mantler:2012bj}).

For most of the channels, the cross-section increase $\rho$ at 100 TeV with respect to the 8 TeV LHC ranges from a factor of 20 to a factor of 50, with the exception of the $HV^\pm jj$ ($V=W^\pm,Z$) final states, where it is around 80, and of the $Ht\bar t$ and $Htj$ final states, where it exceeds 250.
The remarkable growth of $pp\to Htj$, together with its sensitivity to the sign of the top-quark Yukawa coupling $y_t$ \cite{Farina:2012xp}, makes this reaction a golden channel for a precise measurement of the latter. It has been shown \cite{Chang:2014rfa} that already at the 14 TeV LHC it is possible to put loose bounds on the sign of $y_t$, mainly with a semileptonically decaying top quark, and in the $H\to b\bar b$ and $H\to\gamma\gamma$ decay channels. At the FCC-hh the situation will improve considerably, owing to the cross-section and luminosity increase. Indeed, the NLO cross section for the main irreducible background to $tH(\to\gamma\gamma)j$ production, namely $t\gamma\gamma j$ QCD production, has a growth $\rho$ comparable to that of the signal (or even slightly smaller: about 200 times with the photon-isolation setup employed here), hence the significance of the signal, in comparison with the LHC, is expected to scale at least with the square root of the number of events. Moreover, the sensitivity of the signal to the value of the top-quark Yukawa is very similar at the FCC-hh as at the LHC, as shown explicitly in the right panel of figure \ref{fig:H}.

\begin{figure}[h!]
\begin{minipage}{0.49\textwidth}
\centering
\includegraphics[width=1\textwidth]{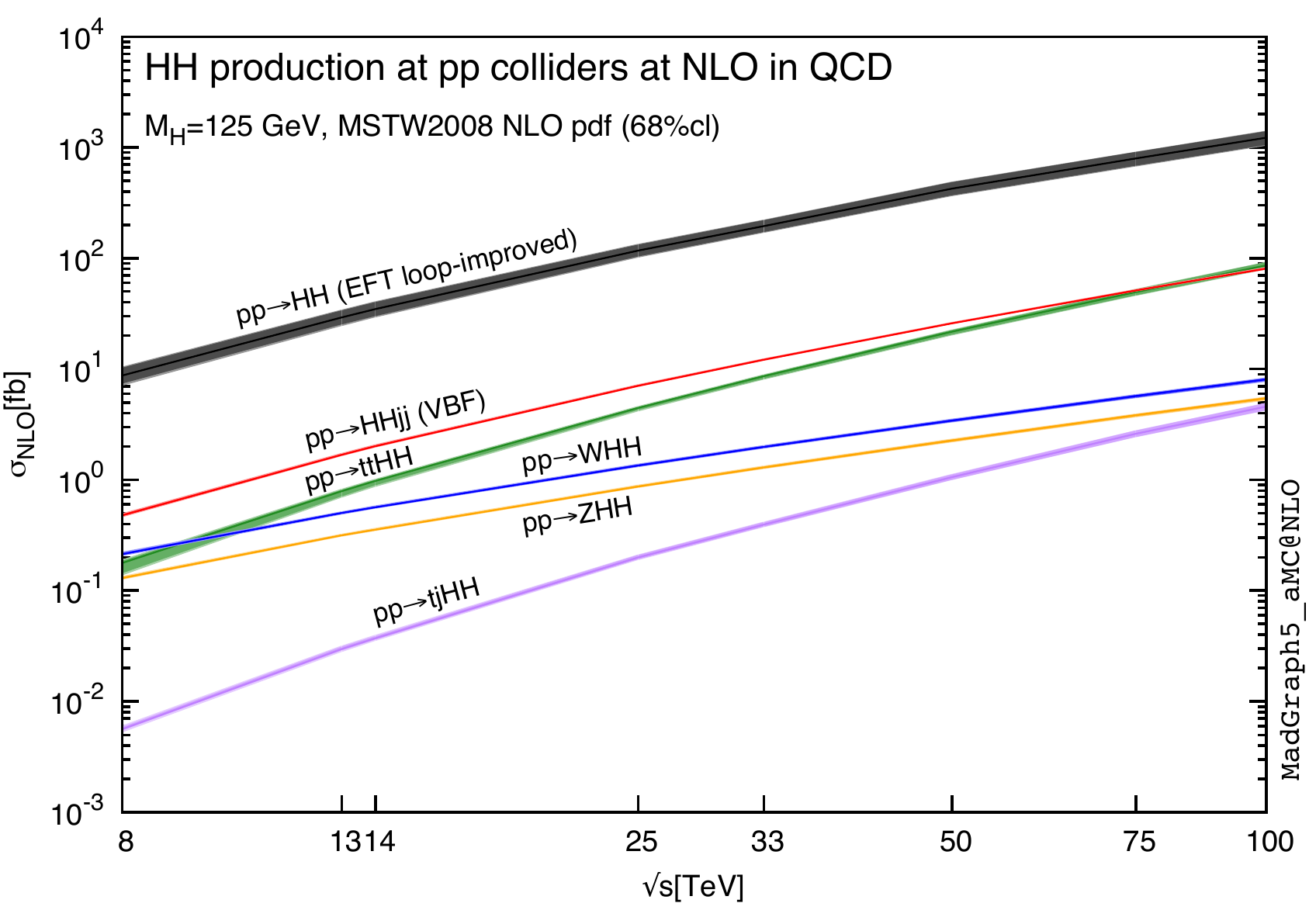}
\end{minipage}
\begin{minipage}{0.49\textwidth}
\centering
\includegraphics[width=1\textwidth]{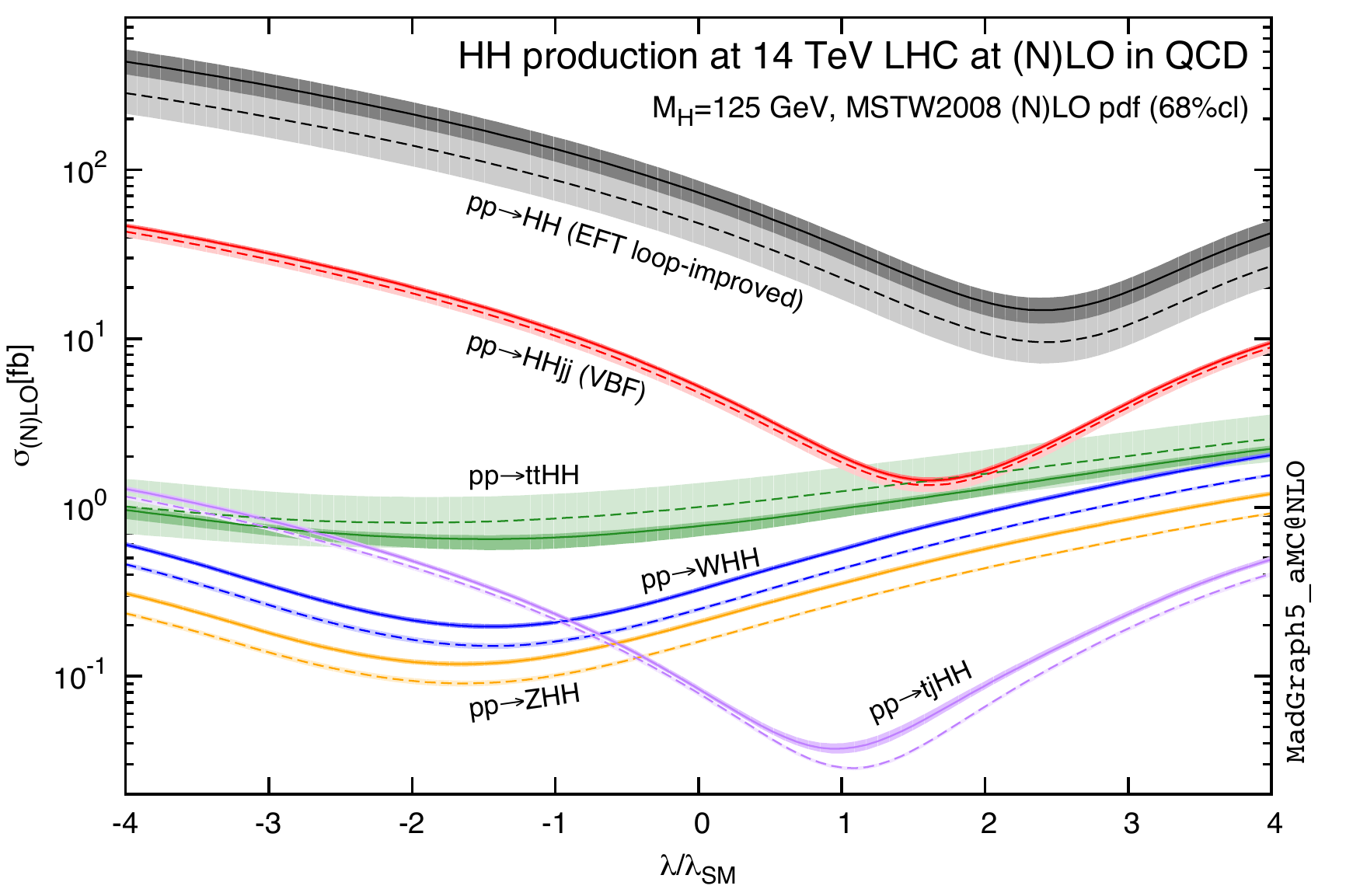}
\end{minipage}
\caption{\label{fig:HH}Left panel: NLO total cross section for Higgs-pair production in association with light and heavy quarks, and electroweak bosons. Right panel: dependence of the LO and NLO total cross section for different Higgs-pair-production channels upon the trilinear Higgs coupling $\lambda$, at the 14 TeV LHC. Plots are taken from \cite{Frederix:2014hta}.}
\end{figure}

The left panel of figure \ref{fig:HH}, taken from \cite{Frederix:2014hta}, shows the dominant Higgs-pair-production channels. The pattern of growth is qualitatively similar to single-Higgs production, with the top-associated channels displaying the largest increase. The importance of this class of processes is linked in particular to their sensitivity to potential deviations from the SM structure of the Higgs self-interactions. The right panel of figure \ref{fig:HH}, also taken from \cite{Frederix:2014hta}, shows interestingly that the dominant mode, namely Higgs-pair production through gluon fusion, is also one of the most sensitive to the variation of the trilinear Higgs coupling $\lambda$.\footnote{The right panel of figure \ref{fig:HH} and the related comments have just to be interpreted as a sketch: a more correct and complete study of non-standard Higgs interactions would take into account all of the operators of a given canonical dimension in an effective-field-theory framework. This is of course beyond the scope of this talk.}

Associated production of three Higgs bosons, relevant to studies of the Higgs quartic self-interaction, can also be simulated in the \amc~framework, and will be documented elsewhere \cite{hhh}.

\begin{table}[h!]
\begin{center}
\begin{small}
\begin{tabular}{rl | l | l | c}
\hline\hline
 &Process& $\sigma_{\tiny{\mbox{NLO}}}$(8 TeV) [fb]& $\sigma_{\tiny{\mbox{NLO}}}$(100 TeV) [fb]\vspace{1mm}&$\rho$\\\hline\vspace{1mm}
$pp~~\to$ & $H~(m_t, m_b)$ & $1.44\cdot 10^4~{}^{+20\%}_{-16\%}~{}^{+1\%}_{-2\%}$ & $5.46\cdot 10^5~{}^{+28\%}_{-27\%}~{}^{+2\%}_{-2\%}$ & 38 \vspace{1mm}\\\hline\vspace{1mm}
$pp~~\to$ & $Hjj~(\mbox{VBF})$ & $1.61\cdot 10^3~{}^{+1\%}_{-0\%}~{}^{+2\%}_{-2\%}$ & $7.40\cdot 10^4~{}^{+3\%}_{-2\%}~{}^{+2\%}_{-1\%}$ & 46 \vspace{1mm}\\\hline\vspace{1mm}
$pp~~\to$ & $Ht\bar t$ & $1.21\cdot 10^2~{}^{+5\%}_{-9\%}~{}^{+3\%}_{-3\%}$ & $3.25\cdot 10^4~{}^{+7\%}_{-8\%}~{}^{+1\%}_{-1\%}$ & 269 \vspace{1mm}\\\hline\vspace{1mm}
$pp~~\to$ & $Hb\bar b~(\mbox{4FS})$ & $2.37\cdot 10^2~{}^{+9\%}_{-9\%}~{}^{+2\%}_{-2\%}$ & $1.21\cdot 10^4~{}^{+2\%}_{-10\%}~{}^{+2\%}_{-2\%}$ & 51 \vspace{1mm}\\\hline\vspace{1mm}
$pp~~\to$ & $Htj$ & $2.07\cdot 10^1~{}^{+2\%}_{-1\%}~{}^{+2\%}_{-2\%}$ & $5.21\cdot 10^3~{}^{+3\%}_{-5\%}~{}^{+1\%}_{-1\%}$ & 252 \vspace{1mm}\\\hline\hline\vspace{1mm}
$pp~~\to$ & $HW^\pm$ & $7.31\cdot 10^2~{}^{+2\%}_{-1\%}~{}^{+2\%}_{-2\%}$ & $1.54\cdot 10^4~{}^{+5\%}_{-8\%}~{}^{+2\%}_{-2\%}$ & 21 \vspace{1mm}\\\hline\vspace{1mm}
$pp~~\to$ & $HZ$ & $3.87\cdot 10^2~{}^{+2\%}_{-1\%}~{}^{+2\%}_{-2\%}$ & $8.82\cdot 10^3~{}^{+4\%}_{-8\%}~{}^{+2\%}_{-2\%}$ & 23 \vspace{1mm}\\\hline\vspace{1mm}
$pp~~\to$ & $HW^+W^-~(\mbox{4FS})$ & $4.62\cdot 10^0~{}^{+3\%}_{-2\%}~{}^{+2\%}_{-2\%}$ & $1.68\cdot 10^2~{}^{+5\%}_{-6\%}~{}^{+2\%}_{-1\%}$ & 36 \vspace{1mm}\\\hline\vspace{1mm}
$pp~~\to$ & $HZW^\pm$ & $2.17\cdot 10^0~{}^{+4\%}_{-4\%}~{}^{+2\%}_{-2\%}$ & $9.94\cdot 10^1~{}^{+6\%}_{-7\%}~{}^{+2\%}_{-1\%}$ & 46 \vspace{1mm}\\\hline\vspace{1mm}
$pp~~\to$ & $HW^\pm\gamma$ & $2.36\cdot 10^0~{}^{+3\%}_{-3\%}~{}^{+2\%}_{-2\%}$ & $7.75\cdot 10^1~{}^{+7\%}_{-8\%}~{}^{+2\%}_{-1\%}$ & 33 \vspace{1mm}\\\hline\vspace{1mm}
$pp~~\to$ & $HZ\gamma$ & $1.54\cdot 10^0~{}^{+3\%}_{-2\%}~{}^{+2\%}_{-2\%}$ & $4.29\cdot 10^1~{}^{+5\%}_{-7\%}~{}^{+2\%}_{-2\%}$ & 28 \vspace{1mm}\\\hline\vspace{1mm}
$pp~~\to$ & $HZZ$ & $1.10\cdot 10^0~{}^{+2\%}_{-2\%}~{}^{+2\%}_{-2\%}$ & $4.20\cdot 10^1~{}^{+4\%}_{-6\%}~{}^{+2\%}_{-1\%}$ & 38 \vspace{1mm}\\\hline\hline\vspace{1mm}
$pp~~\to$ & $HW^\pm j$ & $3.18\cdot 10^2~{}^{+4\%}_{-4\%}~{}^{+2\%}_{-1\%}$ & $1.07\cdot 10^4~{}^{+2\%}_{-7\%}~{}^{+2\%}_{-1\%}$ & 34 \vspace{1mm}\\\hline\vspace{1mm}
$pp~~\to$ & $HW^\pm jj$ & $6.06\cdot 10^1~{}^{+6\%}_{-8\%}~{}^{+1\%}_{-1\%}$ & $4.90\cdot 10^3~{}^{+2\%}_{-6\%}~{}^{+1\%}_{-1\%}$ & 81 \vspace{1mm}\\\hline\vspace{1mm}
$pp~~\to$ & $HZj$ & $1.71\cdot 10^2~{}^{+4\%}_{-4\%}~{}^{+1\%}_{-1\%}$ & $6.31\cdot 10^3~{}^{+2\%}_{-7\%}~{}^{+2\%}_{-1\%}$ & 37 \vspace{1mm}\\\hline\vspace{1mm}
$pp~~\to$ & $HZjj$ & $3.50\cdot 10^1~{}^{+7\%}_{-10\%}~{}^{+1\%}_{-1\%}$ & $2.81\cdot 10^3~{}^{+2\%}_{-5\%}~{}^{+1\%}_{-1\%}$ & 80 \vspace{1mm}\\\hline\hline
\end{tabular}
\end{small}
\end{center}
\caption{\label{tab:H} Production of a single Higgs boson at the LHC and at a 100 TeV FCC-hh. The rightmost column reports the ratio $\rho$ of the FCC-hh to the LHC cross sections. Theoretical uncertainties are due to scale and PDF variations, respectively. Monte-Carlo-integration error is always smaller than theoretical uncertainties, and is not shown. For $pp\to HV jj$, on top of the transverse-momentum cut of section \ref{sec:setup}, I require $m(j_1,j_2)> 100$ GeV, $j_1$ and $j_2$ being the hardest and next-to-hardest jets, respectively. Processes $pp\to Htj$ and $pp\to Hjj$ (VBF) do not feature jet cuts.}
\end{table}


\section{Multi-vector-boson production}
\label{sec:V}
\begin{figure}[h!]
\begin{minipage}{0.49\textwidth}
\centering
\includegraphics[width=1\textwidth]{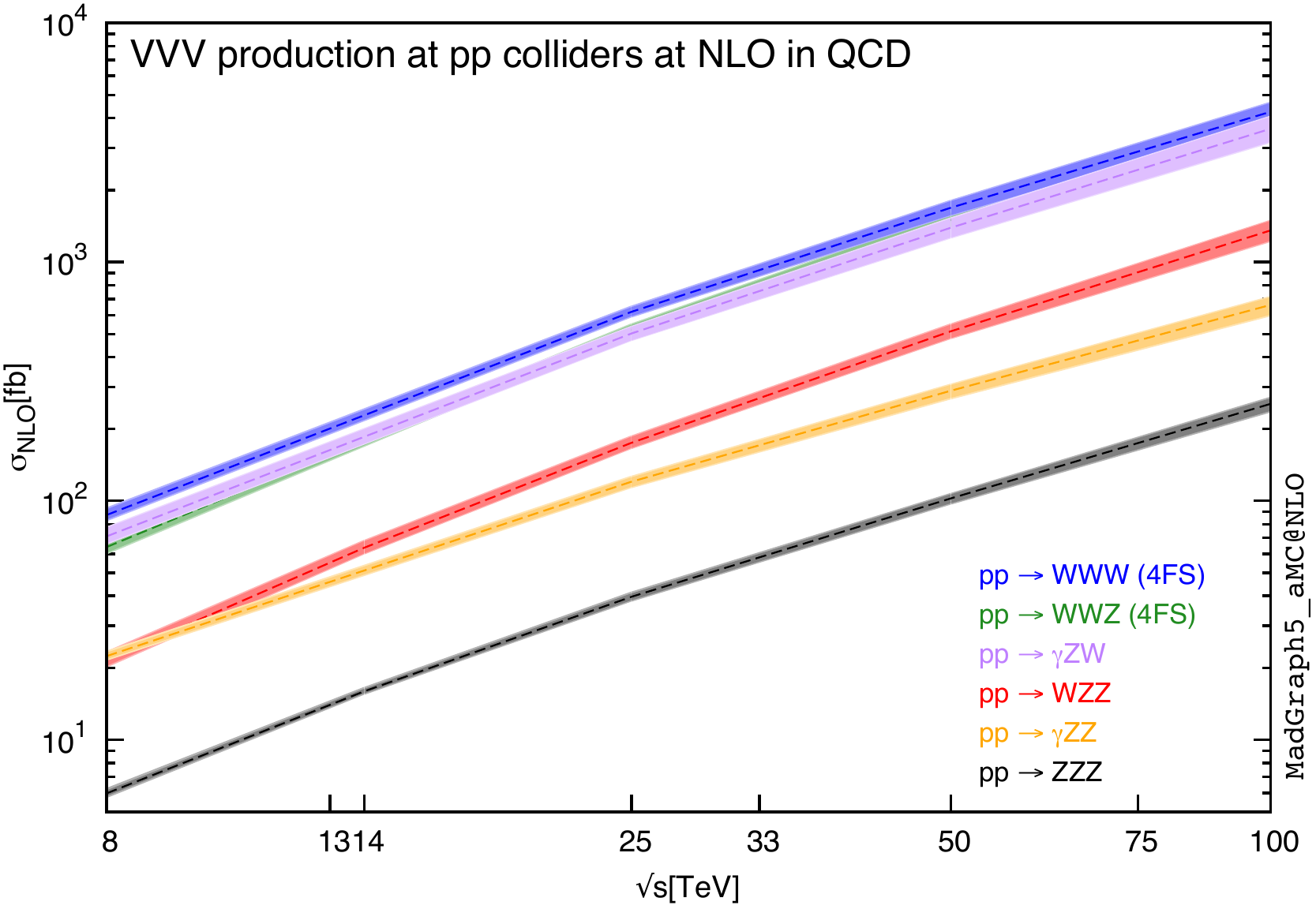}
\end{minipage}
\begin{minipage}{0.49\textwidth}
\centering
\includegraphics[width=1\textwidth]{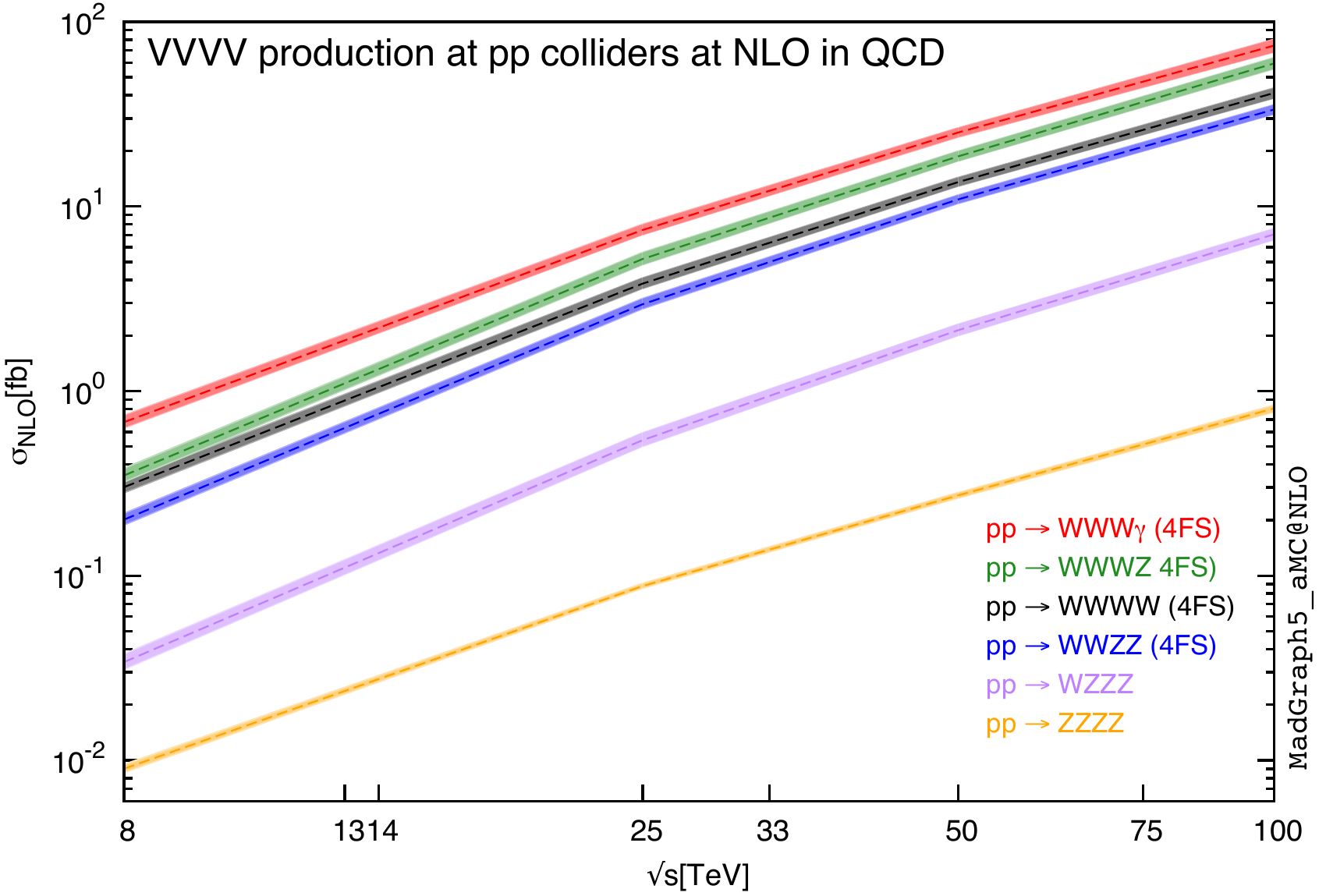}
\end{minipage}
\caption{\label{fig:V}NLO total cross section for production of three (left panel) and four (right panel) electroweak bosons.}
\end{figure}

\begin{figure}[h!]
\begin{minipage}{0.49\textwidth}
\centering
\includegraphics[width=1.02\textwidth]{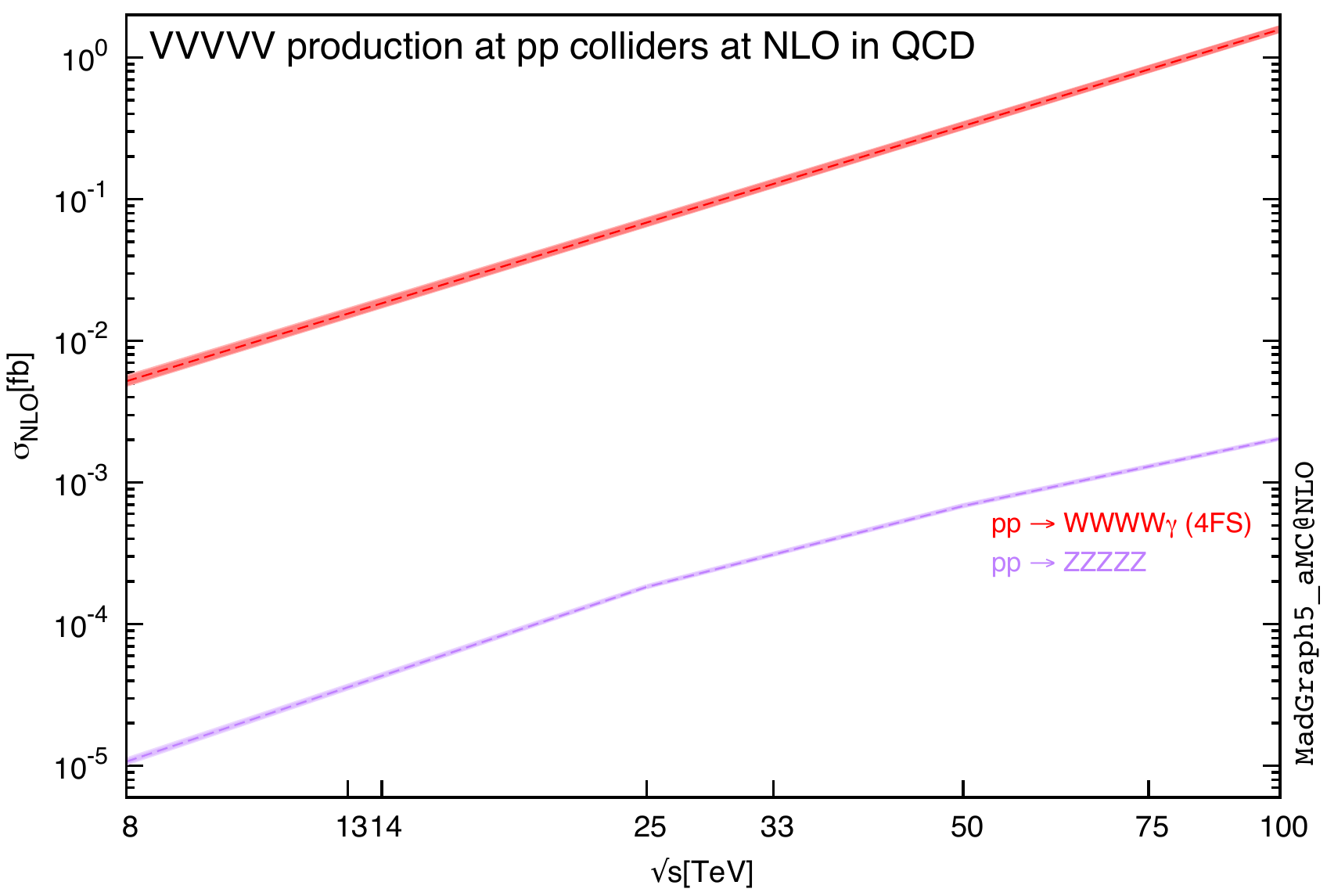}
\end{minipage}
\begin{minipage}{0.49\textwidth}
\centering
\includegraphics[width=1\textwidth]{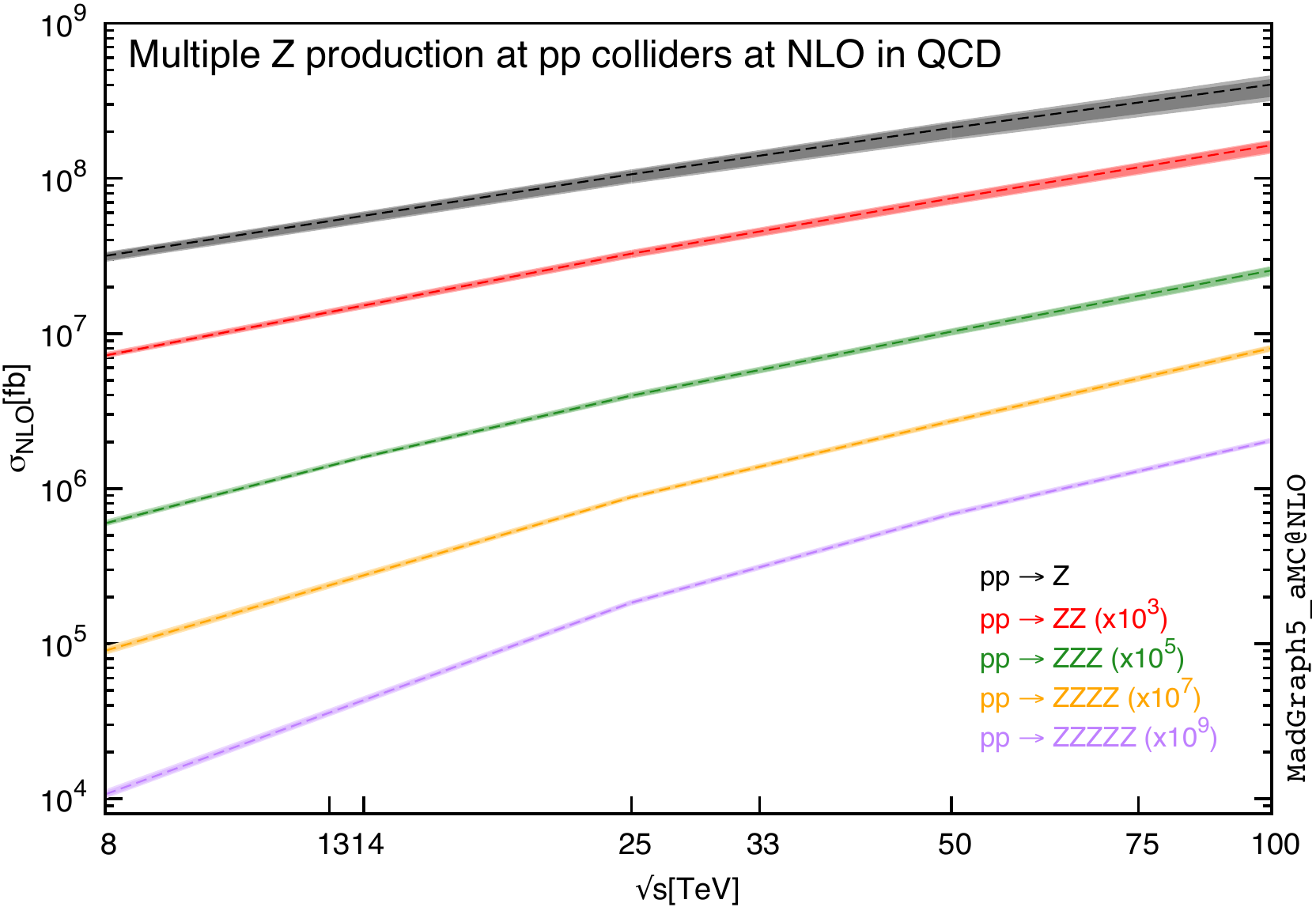}
\end{minipage}
\caption{\label{fig:VV}NLO total cross section for production of five electroweak bosons (left panel) and for $n$ $Z$ bosons, with $n\leq5$ (right panel). Some of the curves in the right panel are rescaled to fit in a single frame. The strictly linear increase in the $pp\to W^+W^-W^+W^-\gamma$ cross section is due to having simulated this process only for 8 and 100 TeV colliders.}
\end{figure}

In table \ref{tab:V} and in figures \ref{fig:V} and \ref{fig:VV}, I report sample cross sections for the production of up to five undecayed electroweak vector bosons. The cross
section for five electroweak bosons is presented here for the first time at the NLO in QCD.
Given the smallness of their production cross section, five-boson final states will be impossible to detect directly at the LHC, and very challenging also at the 100 TeV FCC-hh. In particular, according to the three- and four-boson pattern displayed in table \ref{tab:V} and in the tables of \cite{Alwall:2014hca}, the two five-boson channels simulated here should give a reasonable estimate of the range of cross sections for this class of processes, namely all other five-boson channels (whose simulation is possible, but has been left for future work) should have a total cross section ranging from $\mathcal O(10~\mbox{ab})$ to $\mathcal O(1~\mbox{fb})$ at 100 TeV.

Three-boson cross sections increase by a moderate amount, of the order of 50, while the increase for the four- and the five-boson channels is more significant, and grows with the multiplicity, as displayed in the right panel of figure \ref{fig:VV}, especially at small center-of-mass energies, in the case of many $Z$ bosons. Indeed, at relatively small collider energies, requiring an extra massive particle on top of a heavy system may shrink considerably the available phase space, hence the progressive depletion at small $\sqrt s$ increasing the number of $Z$'s; far from threshold this effect is less pronounced, and the growth for different multiplicities gets more uniform.

Decays of electroweak bosons (as well as those for any other types of resonances) have not been considered in the simulations presented here, but can be included in \amc~with various levels of accuracy. A first approximation consists of letting the bosons stable and then acting \madspin~\cite{Artoisenet:2012st} on the final states. This procedure retains approximate NLO production and decay spin correlations, and off-shell effects of resonant diagrams \cite{Frixione:2007zp}. A complete simulation, which is obviously much more demanding in terms of computational resources, implies the direct generation of processes featuring decay products instead of resonances in the final state. This is by construction exactly NLO accurate, and retains all non-resonant effects as well, but is possible only for smaller multiplicities: so far it has been carried out in the case of up to three leptonically decaying electroweak bosons \cite{Frederix:2011ss,Alwall:2014hca}, and it is probably feasible for four-boson processes as well.

The inclusion of gluon-initiated loop-induced contributions, formally of NNLO accuracy, is also possible (though not yet automatic) in the \amc~environment, as documented for example in \cite{Frederix:2011ss}; even if these terms have not been considered here, their relevance  is in fact expected to increase with the collider centre-of-mass energy, due to the dominance of the gluon luminosity at small Bjorken-$x$.

\begin{table}[h!]
\begin{center}
\begin{small}
\begin{tabular}{rl | l | l | c}
\hline\hline
 &Process& $\sigma_{\tiny{\mbox{NLO}}}$(8 TeV) [fb]& $\sigma_{\tiny{\mbox{NLO}}}$(100 TeV) [fb]\vspace{1mm}&$\rho$\\\hline
$pp~~\to$ & $W^+W^-W^\pm~(\mbox{4FS})$ & $8.73\cdot 10^1~{}^{+6\%}_{-4\%}~{}^{+2\%}_{-2\%}$ & $4.25\cdot 10^3~{}^{+9\%}_{-9\%}~{}^{+1\%}_{-1\%}$ & 49\vspace{1mm}\\\hline\vspace{1mm}
$pp~~\to$ & $W^+W^-Z~(\mbox{4FS})$ & $6.41\cdot 10^1~{}^{+7\%}_{-5\%}~{}^{+2\%}_{-2\%}$ & $4.01\cdot 10^3~{}^{+9\%}_{-9\%}~{}^{+1\%}_{-1\%}$ & 63\vspace{1mm}\\\hline\vspace{1mm}
$pp~~\to$ & $\gamma W^\pm Z$ & $7.11\cdot 10^1~{}^{+8\%}_{-7\%}~{}^{+2\%}_{-1\%}$ & $3.61\cdot 10^3~{}^{+12\%}_{-12\%}~{}^{+1\%}_{-1\%}$ & 51\vspace{1mm}\\\hline\vspace{1mm}
$pp~~\to$ & $W^\pm ZZ$ & $2.16\cdot 10^1~{}^{+7\%}_{-6\%}~{}^{+2\%}_{-2\%}$ & $1.36\cdot 10^3~{}^{+10\%}_{-10\%}~{}^{+1\%}_{-1\%}$ & 63\vspace{1mm}\\\hline\vspace{1mm}
$pp~~\to$ & $\gamma ZZ$ & $2.24\cdot 10^1~{}^{+4\%}_{-3\%}~{}^{+2\%}_{-2\%}$ & $6.62\cdot 10^2~{}^{+8\%}_{-9\%}~{}^{+2\%}_{-1\%}$ & 30\vspace{1mm}\\\hline\vspace{1mm}
$pp~~\to$ & $ZZZ$ & $5.97\cdot 10^0~{}^{+3\%}_{-3\%}~{}^{+2\%}_{-2\%}$ & $2.55\cdot 10^2~{}^{+5\%}_{-7\%}~{}^{+2\%}_{-1\%}$ & 43\vspace{1mm}\\\hline\hline\vspace{1mm}
$pp~~\to$ & $W^+W^-W^\pm\gamma~(\mbox{4FS})$ & $6.78\cdot 10^{-1}~{}^{+8\%}_{-6\%}~{}^{+2\%}_{-2\%}$ & $7.42\cdot 10^1~{}^{+8\%}_{-8\%}~{}^{+1\%}_{-1\%}$ & 109\vspace{1mm}\\\hline\vspace{1mm}
$pp~~\to$ & $W^+W^-W^\pm Z~(\mbox{4FS})$ & $3.48\cdot 10^{-1}~{}^{+8\%}_{-7\%}~{}^{+2\%}_{-2\%}$ & $5.95\cdot 10^1~{}^{+7\%}_{-7\%}~{}^{+1\%}_{-1\%}$ & 171\vspace{1mm}\\\hline\vspace{1mm}
$pp~~\to$ & $W^+W^-W^+W^-~(\mbox{4FS})$ & $3.01\cdot 10^{-1}~{}^{+7\%}_{-6\%}~{}^{+2\%}_{-2\%}$ & $4.11\cdot 10^1~{}^{+7\%}_{-6\%}~{}^{+1\%}_{-1\%}$ & 137\vspace{1mm}\\\hline\vspace{1mm}
$pp~~\to$ & $W^+W^-ZZ~(\mbox{4FS})$ & $2.01\cdot 10^{-1}~{}^{+7\%}_{-6\%}~{}^{+2\%}_{-2\%}$ & $3.34\cdot 10^1~{}^{+6\%}_{-6\%}~{}^{+1\%}_{-1\%}$ & 166\vspace{1mm}\\\hline\vspace{1mm}
$pp~~\to$ & $W^\pm ZZZ$ & $3.40\cdot 10^{-2}~{}^{+10\%}_{-8\%}~{}^{+2\%}_{-2\%}$ & $7.06\cdot 10^0~{}^{+8\%}_{-7\%}~{}^{+1\%}_{-1\%}$ & 208\vspace{1mm}\\\hline\vspace{1mm}
$pp~~\to$ & $ZZZZ$ & $8.72\cdot 10^{-3}~{}^{+4\%}_{-4\%}~{}^{+3\%}_{-2\%}$ & $8.05\cdot 10^{-1}~{}^{+4\%}_{-4\%}~{}^{+2\%}_{-1\%}$ & 92\vspace{1mm}\\\hline\hline\vspace{1mm}
$pp~~\to$ & $W^+W^-W^+W^-\gamma$~(\mbox{4FS}) & $5.18\cdot 10^{-3}~{}^{+8\%}_{-7\%}~{}^{+3\%}_{-2\%}$ & $1.58\cdot 10^{0}~{}^{+6\%}_{-5\%}~{}^{+1\%}_{-1\%}$& 305\vspace{1mm}\\\hline\vspace{1mm}
$pp~~\to$ & $ZZZZZ$ & $1.07\cdot 10^{-5}~{}^{+5\%}_{-4\%}~{}^{+3\%}_{-2\%}$ & $2.04\cdot 10^{-3}~{}^{+3\%}_{-3\%}~{}^{+2\%}_{-1\%}$ & 191\vspace{1mm}\\\hline\hline
\end{tabular}
\end{small}
\end{center}
\caption{\label{tab:V} Production of multiple vector bosons at the LHC and at a 100 TeV FCC-hh. The rightmost column reports the ratio $\rho$ of the FCC-hh to the LHC cross sections. Theoretical uncertainties are due to scale and PDF variations, respectively. Monte-Carlo-integration error is always smaller than theoretical uncertainties, and is not shown.}
\end{table}


\section{Top-antitop associated production}
\label{sec:t}
\begin{figure}[h!]
\begin{minipage}{0.49\textwidth}
\centering
\includegraphics[width=1\textwidth]{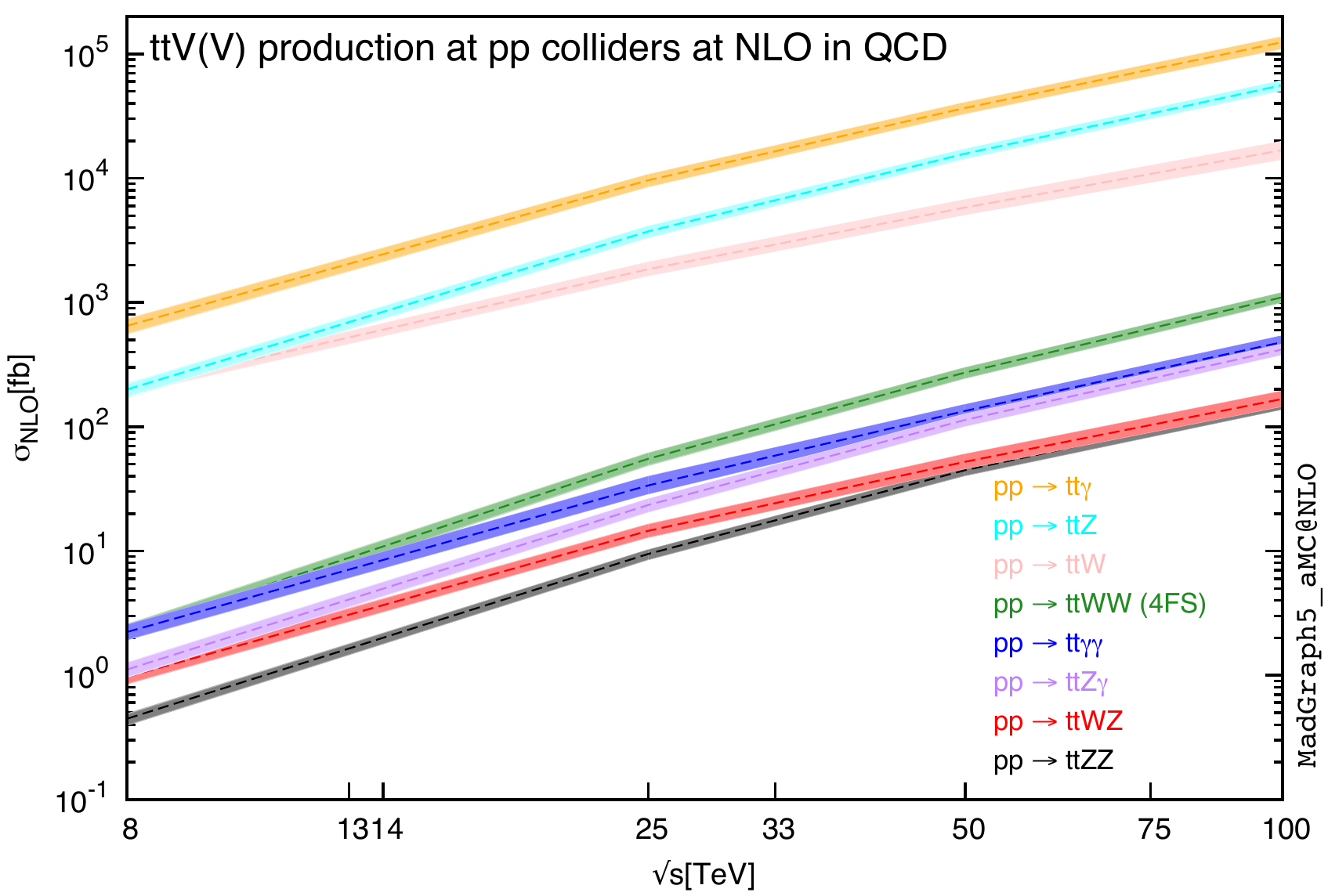}
\end{minipage}
\begin{minipage}{0.49\textwidth}
\centering
\includegraphics[width=1\textwidth]{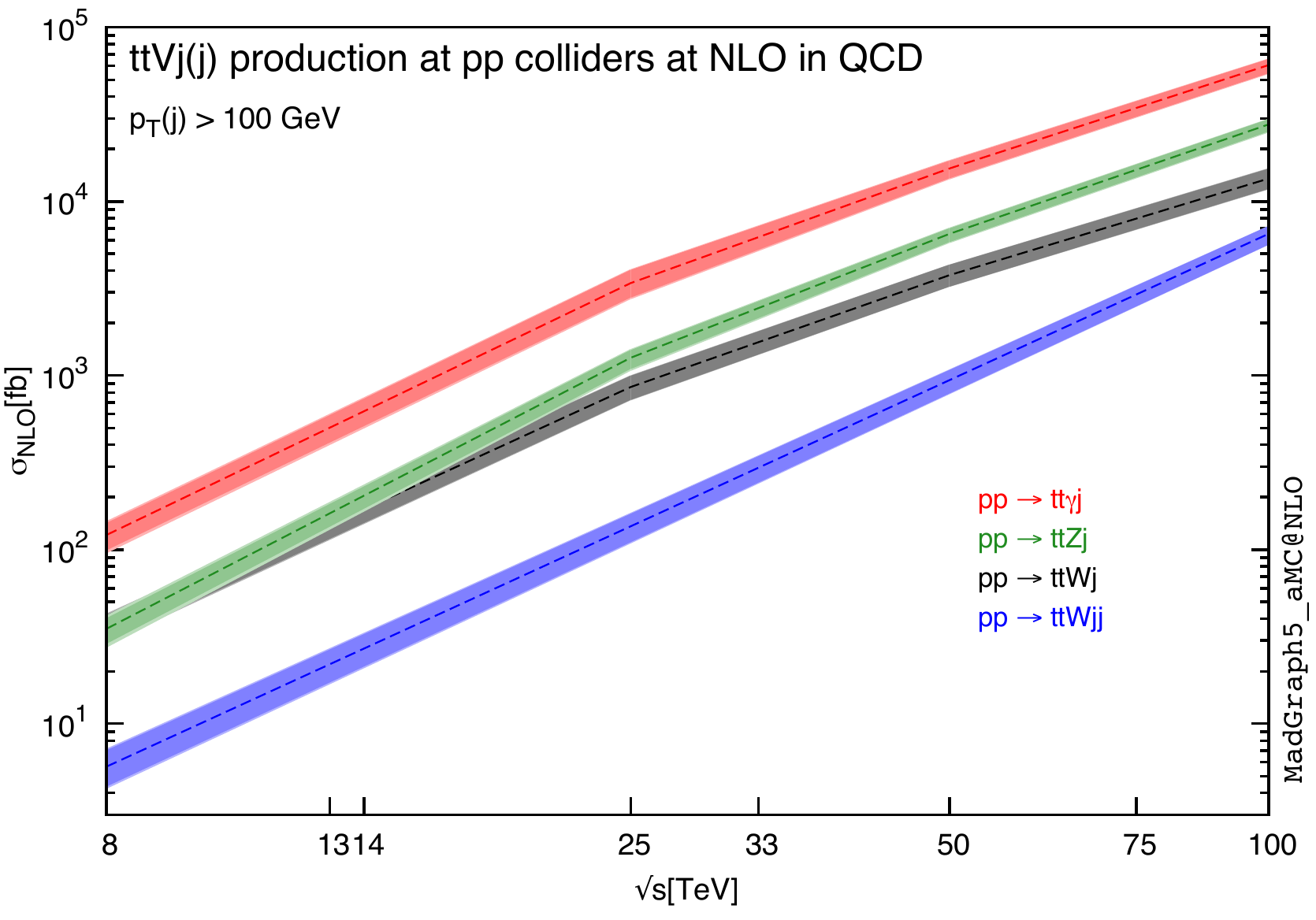}
\end{minipage}
\caption{\label{fig:t}NLO total cross section for production of a top-antitop pair in association with up to two electroweak bosons (left panel), and in association with an electroweak boson and up to two jets (right panel). Jet transverse momenta undergo a cut $p_T(j) > 100$ GeV. The strictly linear increase in the $pp\to t\bar t W^\pm jj$ cross section is due to having simulated this process only for 8 and 100 TeV colliders.}
\end{figure}

In table \ref{tab:t} and in figure \ref{fig:t}, I report results relevant to the production of a top-antitop pair in association with up to two electroweak bosons, and with an electroweak boson and up to two jets. The cross section for a top-antitop pair in association with an electroweak vector boson and two jets is presented here for the first time at the NLO in QCD.

Process $pp\to t\bar t W^\pm$ has been recently studied \cite{Maltoni:2014zpa} in relation to the top-antitop charge asymmetry at proton-proton colliders. The absence of the gluon-fusion channel in this process at LO and NLO is responsible for its more limited cross-section increase $\rho$ with respect to the neutral  $pp\to t\bar tV$ reactions (see table \ref{tab:t}), but is also what enhances the charge asymmetry and makes it relatively stable increasing the centre-of-mass energy, hence making this reaction an interesting handle to constrain New-Physics effects at present and future colliders.

Apart from $pp\to t\bar t W^\pm$, the cross-section increase is quite substantial, of the order of a few hundreds, and generally growing with the final-state multiplicity. The cross section for processes with jets increases by up to three plain orders of magnitudes, which is partly due to the $p_T(j)>100$~GeV threshold that considerably shrinks the available phase space at small collider energies.

As for the channels with two electroweak bosons, it is interesting to notice that the dominant one at the LHC, namely $t\bar tW^+W^-$, is also the one featuring the largest $\rho$. This renders this process well visible at the FCC-hh, with an NLO cross section (before branching ratios) as large as 1 pb. A detailed study of this class of processes will be documented in a dedicated publication \cite{Maltoni}.

Results for two or three top-antitop pairs, obtained in the \amc~framework, have been presented in \cite{Deandrea:2014raa}, and are not reproduced here.

\begin{table}[h!]
\begin{center}
\begin{small}
\begin{tabular}{rl | l | l | c}
\hline\hline
 &Process& $\sigma_{\tiny{\mbox{NLO}}}$(8 TeV) [fb]& $\sigma_{\tiny{\mbox{NLO}}}$(100 TeV) [fb]\vspace{1mm}&$\rho$\\\hline
$pp~~\to$ & $t\bar t\gamma$ & $6.50\cdot 10^2~{}^{+12\%}_{-13\%}~{}^{+2\%}_{-2\%}$ & $1.24\cdot 10^5~{}^{+11\%}_{-11\%}~{}^{+1\%}_{-1\%}$ & 192\vspace{1mm}\\\hline\vspace{1mm}
$pp~~\to$ & $t\bar tZ$ & $1.99\cdot 10^2~{}^{+10\%}_{-12\%}~{}^{+3\%}_{-3\%}$ & $5.63\cdot 10^4~{}^{+9\%}_{-10\%}~{}^{+1\%}_{-1\%}$ & 282\vspace{1mm}\\\hline\vspace{1mm}
$pp~~\to$ & $t\bar tW^\pm$ & $2.05\cdot 10^2~{}^{+9\%}_{-10\%}~{}^{+2\%}_{-2\%}$ & $1.68\cdot 10^4~{}^{+18\%}_{-16\%}~{}^{+1\%}_{-1\%}$ & 82\vspace{1mm}\\\hline\hline\vspace{1mm}
$pp~~\to$ & $t\bar t\gamma j$ & $1.22\cdot 10^2~{}^{+17\%}_{-18\%}~{}^{+3\%}_{-3\%}$ & $6.07\cdot 10^4~{}^{+8\%}_{-10\%}~{}^{+1\%}_{-1\%}$ & 498\vspace{1mm}\\\hline\vspace{1mm}
$pp~~\to$ & $t\bar tZj$ & $3.51\cdot 10^1~{}^{+15\%}_{-18\%}~{}^{+4\%}_{-4\%}$ & $2.77\cdot 10^4~{}^{+7\%}_{-9\%}~{}^{+1\%}_{-1\%}$ & 789\vspace{1mm}\\\hline\vspace{1mm}
$pp~~\to$ & $t\bar tW^\pm j$ & $3.59\cdot 10^1~{}^{+18\%}_{-18\%}~{}^{+2\%}_{-2\%}$ & $1.36\cdot 10^4~{}^{+14\%}_{-13\%}~{}^{+1\%}_{-1\%}$ & 379\vspace{1mm}\\\hline\hline\vspace{1mm}
$pp~~\to$ & $t\bar tW^\pm jj$ & $5.67\cdot 10^0~{}^{+24\%}_{-23\%}~{}^{+3\%}_{-2\%}$ & $6.52\cdot 10^3~{}^{+11\%}_{-14\%}~{}^{+1\%}_{-1\%}$ & 1150\vspace{1mm}\\\hline\hline\vspace{1mm}
$pp~~\to$ & $t\bar tW^+W^-~(\mbox{4FS})$ & $2.27\cdot 10^0~{}^{+11\%}_{-13\%}~{}^{+3\%}_{-3\%}$ & $1.10\cdot 10^3~{}^{+9\%}_{-9\%}~{}^{+1\%}_{-1\%}$ & 486\vspace{1mm}\\\hline\vspace{1mm}
$pp~~\to$ & $t\bar t\gamma\gamma$ & $2.23\cdot 10^0~{}^{+14\%}_{-13\%}~{}^{+2\%}_{-1\%}$ & $4.81\cdot 10^2~{}^{+13\%}_{-11\%}~{}^{+1\%}_{-1\%}$ & 216\vspace{1mm}\\\hline\vspace{1mm}
$pp~~\to$ & $t\bar tZ\gamma$ & $1.11\cdot 10^0~{}^{+12\%}_{-13\%}~{}^{+2\%}_{-2\%}$ & $4.20\cdot 10^2~{}^{+10\%}_{-9\%}~{}^{+1\%}_{-1\%}$ & 378\vspace{1mm}\\\hline\vspace{1mm}
$pp~~\to$ & $t\bar tW^\pm Z$ & $9.71\cdot 10^{-1}~{}^{+10\%}_{-11\%}~{}^{+3\%}_{-2\%}$ & $1.68\cdot 10^2
~{}^{+16\%}_{-13\%}~{}^{+1\%}_{-1\%}$ & 173\vspace{1mm}\\\hline\vspace{1mm}
$pp~~\to$ & $t\bar tZZ$ & $4.47\cdot 10^{-1}~{}^{+8\%}_{-10\%}~{}^{+3\%}_{-2\%}$ & $1.58\cdot 10^2~{}^{+15\%}_{-12\%}~{}^{+1\%}_{-1\%}$ & 353\vspace{1mm}\\\hline\hline
\end{tabular}
\end{small}
\end{center}
\caption{\label{tab:t} Production of a top-antitop pair in association with up to two electroweak vector bosons, and with an electroweak boson and up to two jets, at the LHC and at a 100 TeV FCC-hh. The rightmost column reports the ratio $\rho$ of the FCC-hh to the LHC cross sections. Processes $pp\to t\bar tVj(j)$ feature a cut of $p_T(j)>100$ GeV. Theoretical uncertainties are due to scale and PDF variations, respectively. Monte-Carlo-integration error is always smaller than theoretical uncertainties, and is not shown.}
\end{table}

\section{Conclusions}
I have presented the scaling of the total cross section for various complex Standard-Model processes involving many Higgs bosons, electroweak bosons, or top quarks in the final state, with the aim of assessing the magnitude of these rare reactions at present and future colliders, in view of the preliminary physics studies of the future circular hadronic collider. Results for five electroweak-boson production and for the production of a top-antitop pair with an electroweak boson and two jets have been presented here for the first time at the NLO in QCD. All predictions have been obtained automatically in the \amc~framework.

\label{sec:conc}
\section*{Acknowledgements}
I am grateful to S.~Frixione, F.~Maltoni, T.~Gehrmann, and A.~Papaefstathiou for helpful discussions, and to V.~Hirschi and M.~Zaro for technical support. This research is supported in part by the Swiss National Science Foundation (SNF) under contract 200020-149517 and by the European Commission through the``LHCPhenoNet" Initial Training Network PITN-GA-2010-264564.


\begin{thebibliography}{100}

\bibitem{Aad:2012tfa}
  G.~Aad {\it et al.}  [ATLAS Collaboration],
  Phys.\ Lett.\ B {\bf 716} (2012) 1
  [arXiv:1207.7214 [hep-ex]].

\bibitem{Chatrchyan:2012ufa}
  S.~Chatrchyan {\it et al.}  [CMS Collaboration],
  Phys.\ Lett.\ B {\bf 716} (2012) 30
  [arXiv:1207.7235 [hep-ex]].
  
\bibitem{Alwall:2014hca}
  J.~Alwall, R.~Frederix, S.~Frixione, V.~Hirschi, F.~Maltoni, O.~Mattelaer, H.~-S.~Shao and T.~Stelzer {\it et al.},
  arXiv:1405.0301 [hep-ph].

\bibitem{Alwall:2011uj}
  J.~Alwall, M.~Herquet, F.~Maltoni, O.~Mattelaer and T.~Stelzer,
  JHEP {\bf 1106} (2011) 128
  [arXiv:1106.0522 [hep-ph]].

\bibitem{Frixione:1995ms}
  S.~Frixione, Z.~Kunszt and A.~Signer,
  Nucl.\ Phys.\ B {\bf 467} (1996) 399
  [hep-ph/9512328].

\bibitem{Frederix:2009yq}
  R.~Frederix, S.~Frixione, F.~Maltoni and T.~Stelzer,
  JHEP {\bf 0910} (2009) 003
  [arXiv:0908.4272 [hep-ph]].

\bibitem{Hirschi:2011pa}
  V.~Hirschi, R.~Frederix, S.~Frixione, M.~V.~Garzelli, F.~Maltoni and R.~Pittau,
  JHEP {\bf 1105} (2011) 044
  [arXiv:1103.0621 [hep-ph]].
  
\bibitem{Ossola:2006us}
  G.~Ossola, C.~G.~Papadopoulos and R.~Pittau,
  Nucl.\ Phys.\ B {\bf 763} (2007) 147
  [hep-ph/0609007].

\bibitem{Ossola:2007ax}
  G.~Ossola, C.~G.~Papadopoulos and R.~Pittau,
  JHEP {\bf 0803} (2008) 042
  [arXiv:0711.3596 [hep-ph]].

\bibitem{Cascioli:2011va}
  F.~Cascioli, P.~Maierhofer and S.~Pozzorini,
  Phys.\ Rev.\ Lett.\  {\bf 108} (2012) 111601
  [arXiv:1111.5206 [hep-ph]].
   
\bibitem{Frixione:2002ik}
  S.~Frixione and B.~R.~Webber,
  JHEP {\bf 0206} (2002) 029
  [hep-ph/0204244].

\bibitem{Torrielli:2010aw}
  P.~Torrielli and S.~Frixione,
  JHEP {\bf 1004} (2010) 110
  [arXiv:1002.4293 [hep-ph]].

\bibitem{Frixione:2010ra}
  S.~Frixione, F.~Stoeckli, P.~Torrielli and B.~R.~Webber,
  JHEP {\bf 1101} (2011) 053
  [arXiv:1010.0568 [hep-ph]].

\bibitem{PY8unp}
  S.~Prestel and P.~Torrielli,
  unpublished.

\bibitem{Frederix:2012ps}
  R.~Frederix and S.~Frixione,
  JHEP {\bf 1212} (2012) 061
  [arXiv:1209.6215 [hep-ph]].

\bibitem{Frederix:2011ss}
  R.~Frederix, S.~Frixione, V.~Hirschi, F.~Maltoni, R.~Pittau and P.~Torrielli,
  JHEP {\bf 1202} (2012) 099
  [arXiv:1110.4738 [hep-ph]].

\bibitem{Martin:2009iq}
  A.~D.~Martin, W.~J.~Stirling, R.~S.~Thorne and G.~Watt,
  Eur.\ Phys.\ J.\ C {\bf 63} (2009) 189
  [arXiv:0901.0002 [hep-ph]].
  
\bibitem{Cacciari:2008gp}
  M.~Cacciari, G.~P.~Salam and G.~Soyez,
  JHEP {\bf 0804} (2008) 063
  [arXiv:0802.1189 [hep-ph]].

\bibitem{Cacciari:2011ma}
  M.~Cacciari, G.~P.~Salam and G.~Soyez,
  Eur.\ Phys.\ J.\ C {\bf 72} (2012) 1896
  [arXiv:1111.6097 [hep-ph]].

\bibitem{Frixione:1998jh}
  S.~Frixione,
  Phys.\ Lett.\ B {\bf 429} (1998) 369
  [hep-ph/9801442].

\bibitem{Harlander:2012pb}
  R.~V.~Harlander, S.~Liebler and H.~Mantler,
  Computer Physics Communications {\bf 184} (2013) 1605
  [arXiv:1212.3249 [hep-ph]].

\bibitem{Mantler:2012bj}
  H.~Mantler and M.~Wiesemann,
  Eur.\ Phys.\ J.\ C {\bf 73} (2013) 2467
  [arXiv:1210.8263 [hep-ph]].

\bibitem{Farina:2012xp}
  M.~Farina, C.~Grojean, F.~Maltoni, E.~Salvioni and A.~Thamm,
  JHEP {\bf 1305} (2013) 022
  [arXiv:1211.3736 [hep-ph]].

\bibitem{Chang:2014rfa}
  J.~Chang, K.~Cheung, J.~S.~Lee and C.~-T.~Lu,
  JHEP {\bf 1405} (2014) 062
  [arXiv:1403.2053 [hep-ph]].

\bibitem{Frederix:2014hta}
  R.~Frederix, S.~Frixione, V.~Hirschi, F.~Maltoni, O.~Mattelaer, P.~Torrielli, E.~Vryonidou and M.~Zaro,
  Phys.\ Lett.\ B {\bf 732} (2014) 142
  [arXiv:1401.7340 [hep-ph]].

\bibitem{hhh}
  F.~Maltoni, E.~Vryonidou, and M.~Zaro,
  in preparation.

\bibitem{Artoisenet:2012st}
  P.~Artoisenet, R.~Frederix, O.~Mattelaer and R.~Rietkerk,
  JHEP {\bf 1303} (2013) 015
  [arXiv:1212.3460 [hep-ph]].
  
\bibitem{Frixione:2007zp}
  S.~Frixione, E.~Laenen, P.~Motylinski and B.~R.~Webber,
  JHEP {\bf 0704} (2007) 081
  [hep-ph/0702198 [HEP-PH]].

\bibitem{Maltoni:2014zpa}
  F.~Maltoni, M.~L.~Mangano, I.~Tsinikos and M.~Zaro,
  arXiv:1406.3262 [hep-ph].

\bibitem{Maltoni}
  F.~Maltoni, D.~Pagani and I.~Tsikonos,
  in preparation.

\bibitem{Deandrea:2014raa}
  A.~Deandrea and N.~Deutschmann,
  arXiv:1405.6119 [hep-ph].


  
\end{thebibliography}
\end{document}